\documentclass[12pt]{aastex}
\begin{document}

\title{Coronal Mass Ejections As a Mechanism for Producing IR~Variability in Debris Disks}
\author{Rachel Osten,\altaffilmark{1} Mario Livio,\altaffilmark{1} Steve Lubow,\altaffilmark{1}  J.~E.\ Pringle,\altaffilmark{1,2}\\ David Soderblom,\altaffilmark{1} and Jeff Valenti\altaffilmark{1}}

\altaffiltext{1}{Space Telescope Science Institute, Baltimore, MD  21218}
\altaffiltext{2}{Institute of Astronomy, University of Cambridge, Madingley Road, Cambridge CB3 0HA}

\begin{abstract}
Motivated by recent observations of short-timescale variations in the infrared emission of circumstellar disks, we propose that coronal mass ejections can remove dust grains on timescales as short as a few days.  Continuous monitoring of stellar activity, coupled to infrared observations, can place meaningful constraints on the proposed mechanism.
\end{abstract}

\section{Introduction}

Two recent papers show that the radiation associated with warm, dusty circumstellar disks around
cool stars can vary on surprisingly short timescales.

Melis et~al.\ (2012) report on observations of the star TYC 8241 2652~1 taken at around 10~$\mu$m and 20~$\mu$m in which the flux dropped by a factor of about 30 between 2008 May 6 and 2010 Jan 8--10, a period of less than two years. TYC 8241 2652~1 (hereafter TYC 8241) is thought to be a star of approximate spectral type K2, and an age $\sim$10 Myr, situated at a distance of about 140~pc. 

Meng et~al.\ (2010) report on 24~$\mu$m variability in two solar-like stars ID8 and HD~23514. In ID8 the flux was observed to increase by about 30~percent over a two year period, and to then decrease by a similar amount over about a year. In HD~23514 the flux was observed to drop by about 10~percent over a period of a few months, or less. ID8 is a G6--G7 dwarf at a distance of $\sim$450~pc, and HD~23514 is of spectral type F5V at a distance of $\sim$130~pc (e.g., Soderblom et~al.\ 1993). Age estimates for both stars are in the range 30--130 Myr.

Motivated by these observations, we examine in the current paper potential mechanisms for obtaining short-timescale variability.  In particular, we propose that coronal mass ejections (CMEs) may play a hitherto unrecognized role in producing variability in the IR excess of debris disks.

In Section~2 we describe the observations and some of the suggested interpretations, and we discuss CMEs and their potential effects. In Section~3 we briefly discuss dust grain properties, and our conclusions follow.

\section{The IR Fluxes and Variability}

Melis et~al.\ (2012) model the pre-2009 IR excess with emission from an optically thin collection of grains of size 0.3~$\mu$m and a temperature of 450~K, orbiting TYC 8241 at a distance of $\approx$0.4~AU. They find that the ratio of IR excess to stellar luminosity is $L_\mathrm{IR}/L_\ast \approx 0.11$. They also note that given the luminosity of TYC~8241 ($L_\ast \approx 0.7~L_\odot$), grains with radii of $\sim$0.2~$\mu$m or smaller would be radiatively ejected from the system. Based on these parameters, they estimate that roughly $5 \times 10^{21}$~g in grains with radii of order 0.3~$\mu$m are required to produce the pre-2009 epoch IR excess. 

Similar values apply to the stars reported by Meng et~al.\ (2010), except that in ID8 and HD~23514 the fractional IR luminosities are lower ($L_\mathrm{IR}/L_\ast \approx 0.02$), and the stars are more luminous, implying that the dust has to be placed somewhat farther away, at around $\sim$1~AU. Meng et~al.\ (2010) also note that the spectra of HD 23514 and ID8 (and of a few other extreme excess stars) indicate the presence of fine crystalline silicates or silica dust, requiring grain sizes of order $0.1~\mu$m. This is well below the blowout size (see below) for these stars.

The grain blowout size due to radiation pressure is given by (see e.g., Chen \& Jura 2001) 
\begin{equation}
a < \frac{3 L_\ast Q}{16 \pi M_\ast \rho c}
\end{equation}
where $L_\ast$ and $M_\ast$ are the stellar luminosity and mass, $Q$ is the radiation pressure coupling coefficient, $\rho$ is the density of an individual grain, and $c$ is the speed of light.  Grains whose radii satisfy condition~(1) are expelled by radiation. For grains much larger than the mean photon wavelength (as expected for the stars under discussion), the effective cross section can be approximated by the geometrical cross section, so that $Q \approx 1$. Consequently, for grain density $\rho \approx 2$~g~cm$^{-3}$ we find for solar parameters that $a_\mathrm{min} \approx 0.3~\mu$m. 

For the types of stars we are considering here, the assumed grain sizes are close to the blowout limit. Consequently, we can assume that the IR emission is associated with the smallest grains that can survive being blown out.

\subsection{Conditions for Increasing the Flux} 
Only one of the three stars that exhibits short-timescale mid-IR variability, ID8, shows an unambiguous \textit{increase} in the flux (by about 30\% over a period of less than two years; Meng et~al.\ 2010).  This increase, however, may not require the occurrence of any particular event, but rather it may simply represent the replenishment of the small-scale end of a standard cascade distribution in the sizes of particles (of the type ($n(D) \propto D^{-3.5}$).  In other words, if the disk were to start with the entire collisional cascade in place (with a distribution of sizes from dust to large bodies), then the IR flux would decrease if a mechanism were found to remove dust grains of sizes $\sim$1~$\mu$m or smaller.  Following such a decrease, however, the timescale to regenerate the observationally inferred mass of $\sim$0.1~$\mu$m particles by collisional cascade is only of order 0.1~yr---much shorter than the observed timescale for the flux increase (Meng et~al.\ 2010; Melis et~al.\ 2012).

Consequently, we will concentrate in the present work on identifying a mechanism that can \textit{decrease} the flux on a timescale of less than a year, by way of removal of dust at the short end of the size distribution.

\subsection{A Mechanism for Decreasing the Flux}
Melis et~al.\ (2012) considered the possibility of vaporizing the small grain population using the high-energy flux from a stellar flare. They found, however, that x-rays are rather ineffective at heating grains much larger than 0.3~$\mu$m.  Furthermore, they showed that to obtain the necessary vaporization would have required a total x-ray energy of $E\sim 10^{39}$~ ergs, and an x-ray luminosity of $10^2$--$10^3$ times the stellar luminosity.  Melis et~al.\ (2012) noted that flares of such magnitudes have never been observed.

Here we would like to examine the potential effects of coronal mass ejections (CMEs), which for the Sun, occur together with 
solar flares in solar eruptive events.  These are occasions when magnetic reconnection causes a reconfiguration of the magnetic field and liberation of energy, which goes into mass motions, particle acceleration and plasma heating.  First, we note that the observed total kinetic energy released in solar CMEs is comparable to the radiated flare energy (Emslie et al. 2005, 2012).
This is consistent with theoretical expectations from models of magnetic reconnection through magnetohydrodynamic processes. The latter suggest that energy is released equally into ohmic heating and into fluid motions (e.g., Priest \& Forbes 2000). 
About 25\% of the bolometric flare energy in solar flares arises in the form of UV and soft x-ray radiation (Benz 2008).   

We can estimate how large a CME would be required to remove the grains that contribute to the IR excess, by taking the CME to consist of an ionized plasma which is coupled to a magnetic field. As long as the dust grains are charged and have sufficiently small gyroradii  (see discussion in Section~3), they would be swept up by such a magnetic plasma.
The dust grains would be charged by the ionizing radiation of the flare that accompanies the CME.

The ejection velocity of the mass in the CME, $V_\mathrm{CME}$, can be expected to be of the order of the escape velocity from the central star, therefore
\begin{equation}
V_\mathrm{CME} = \left( \frac{2GM_\ast}{R_\ast} \right)^{1/2} = 6.2 \times 10^7~{\rm cm~s}^{-1} 
\left( \frac {M_\ast}{M_{\odot}}\right)^{1/2}
\left(\frac {R_\ast}{R_{\odot}}\right)^{-1/2}~~.
\end{equation}
The velocity distribution of solar CMEs is actually quite broad, ranging from a few tens of km s$^{-1}$ to 3000 km~s$^{-1}$
(Chen et al. 2006).
While the velocity could be somewhat reduced (e.g. by interplanetary draf; Gopalswamy et al. 2006), there
is a wide range of values that would give short timescales for dust removal (see equation (4) below).

If we assume that the mass of dust that needs to be expelled is $m_\mathrm{dust} \simeq 10^{21}$~g, and that the dust is located at a radius of $R_\mathrm{dust} \simeq1$~AU, then from the conservation of momentum we need the mass in the CME, $m_\mathrm{CME}$, to satisfy 
\begin{equation}
m_\mathrm{CME} = m_\mathrm{dust} \frac{V_e^\mathrm{dust}}{V_\mathrm{CME}}\simeq 0.07 m_\mathrm{dust}~~.
\end{equation}
where $V_e^\mathrm{dust}$ is the escape velocity of the dust (at $R_\mathrm{dust}$).  Thus, we require $m_\mathrm{CME}\sim10^{20}$~g in order to expel a mass of $m_\mathrm{dust}\sim10^{21}$~g. We note that while CMEs are somewhat directional, the distribution of their opening angles extends all the way to about 160 degrees, with more energetic CMEs having larger opening angles (e.g., Gopalswamy et~al.\ 2009; Aarnio et~al.\ 2011).  
While CMEs have not been conclusively detected on stars other than the Sun (Leitzinger et al. 2011), the largest solar flares
have a nearly 100\% association with large CMEs (Yashiro et al. 2006).
The precise relation between $m_\mathrm{CME}$ and the flare energy is not known, but estimates based on equipartition between flare-radiated energy and CME kinetic energy (e.g., Emslie et~al.\ 2005, 2012; Osten et~al.\ in prep.) give (for $m_\mathrm{CME}\sim10^{20}$~g) a required flare energy of (a few) $\times10^{35}$~ergs, within
the realm of previously observed large stellar flares.
This calculation envisages a single flare/CME, but 
in principle, several smaller flare/CME pairs occurring within a short period of time ($\lesssim$ 10$^{5}$ s) could also do the job.
 
The timescale on which the dust can be removed is given approximately by
\begin{equation}
\tau_\mathrm{removal} \simeq \frac{R_\mathrm{dust}}{V_\mathrm{CME}} = 2.4 \times 10^5~{\rm s} 
\left(\frac{R_\mathrm{dust}}{1~{\rm AU}}\right)
\left(\frac{M_\ast}{M_{\odot}}\right)^{-1/2}
\left(\frac{R_\ast}{R_{\odot}}\right)^{1/2}~~,
\end{equation}
which leads, for the values indicated, to timescales of a few days.
\section{Grain Properties}

In order for a CME (an ionized magnetic plasma) to be able to sweep up the dust grains, those grains need: (i)~to be charged, and (ii)~their gyroradii need to be sufficiently small, $r_\mathrm{gyro} < R_\mathrm{dust}$, for them to be able to react to the flowing CME.  
Otherwise, any substructure in the CME should not affect the process in any significant way.
The first condition appears to be easily satisfied.  Since the work function for grains is of order 5 eV (e.g., Tielens 2010), we need to consider the number of photons with energies $\gtrsim10$ eV, i.e. the FUV and XEUV wavelength ranges.  In a flare with an energy of $E_\mathrm{flare}\sim10^{35}$~ergs, one can expect the number of such photons to be $N_\mathrm{ph}\sim6\times10^{44}$.  At the same time, the number of grains is of order $N_g=m_\mathrm{dust}/m_g\sim10^{34}$ (where $m_g$ is a typical mass of a grain).  Since the electron yield per photon is of order $l_e/l_a\sim0.1$ (Tielens 2010), where $l_a$ is a typical photon absorption length and $l_e$ is a typical electron collision length, and $N_{ph}\gg $N$_{g}$, we can safely assume that the net charge per grain, $Z$, satisfies $Z\gg1$.

The gyroradius of a grain is given by 
\begin{equation}
r_\mathrm{gyro}=\frac{c}{Ze} \frac{m_g V_e^\mathrm{dust}}{B}\simeq
4~\mathrm{AU} \left(\frac{Z}{1}\right)^{-1}
\left(\frac{m_g}{2.3\times10^{-13}g}\right)
\left(\frac{V_e^\mathrm{dust}}{4.2\times10^6~\mathrm{cm~s}^{-1}}\right)
\left(\frac{B}{G}\right)^{-1}~~,
\end{equation}
where $c$ is the speed of light, $e$ is the electron charge, $B$ is the magnetic field strength (at 1~AU), and the mass of the grain $m_g$ is scaled with the value for a grain of radius 0.3~$\mu$m.  While magnetic fields have not been measured in stellar flares or CMEs, we can estimate the field in the CME by noting that the energy of the accompanying flare is expected to scale with the magnetic energy density in the flaring volume, $E\propto B^2$ (Schrijver 2011).  Using solar flares as reference, and
assuming a dipole field configuration, we would thus obtain $(B/10^{-3}G) \sim(E_\mathrm{flare}/10^{28}~\mathrm{erg})^{1/2}$, which for $E\sim10^{35}$~ergs would yield $B\sim$ a few G. 
We note here that Gopalswamy (2006)  found solar CMEs at 1 AU to have an average value of $\sim$10$^{-4}$ G, within a factor of
ten of the value obtained by the above scaling. 
Remembering that we also expect $Z\gg1$, equation~(5) therefore indicates that 
even with a potential reduction in the field value due to opening angle considerations or
reduction in field strength, $r_\mathrm{gyro}<R_\mathrm{dust}$, as required for dust removal by CMEs.

\section{Summary and Conclusions}

We have shown that CMEs could remove IR-emitting dust on timescales of days.  This of course requires the stars in question to be active, and to have significant flares. We note that HD~23514, one of the stars that motivated this research, has not been detected in an x-ray survey of the Pleiades (Daniel, Linsky, \& Gagn\'e 2002).  However, one star that has been detected (HII 1122) actually shows lower rotational velocity than HD~23514.  Hence, it is not implausible that the latter is an active star.

More importantly, the present work highlights the need for continuous monitoring of these and similar stars, to be able to correctly assess their activity level, in conjunction with IR flux variations.  Furthermore, active M~stars with debris disks would provide for ideal targets to test the ideas presented in this paper.  The currently existing data on transient mass loss that is associated with stellar activity (e.g., Leitzinger et~al.\ 2011) is rather insufficient for placing meaningful observational constraints.  Another promising avenue could be to observe the radio signatures of MHD shocks that are expected to be produced as CMEs traverse the outer stellar atmospheres at Super-Alfvenic speeds.  These would be the stellar equivalents of solar "type II" bursts (e.g., Gopalswamy et~al.\ 2009).

\end{document}